\documentstyle[pra,aps]{revtex}
\newcommand{\BE}{\begin{equation}}
\newcommand{\EE}{\end{equation}}
\newcommand{\BEA}{\begin{eqnarray}}
\newcommand{\EEA}{\end{eqnarray}}
\newcommand{\BEAN}{\begin{eqnarray*}}
\newcommand{\EEAN}{\end{eqnarray*}}
\begin{document}
\draft

\title{A new type of irregular motion in a class of game dynamics systems}

\author{Tsuyoshi Chawanya
\footnote{E-mail address: chawanya@yukawa.kyoto-u.ac.jp}
}
\address{Yukawa Institute for Theoretical Physics, Kyoto 606, Japan}
\date{February 23, 1995}
\maketitle

\begin{abstract}

A new type of asymptotic behavior in a game dynamics system is
discovered. The system exhibits behavior which combines chaotic motion
and attraction to heteroclinic cycles; the trajectory visits several
unstable stationary states repeatedly with an irregular order, and the
typical length of the stay near the steady states grows exponentially
with the number of visits. The dynamics underlying this irregular
motion is analyzed by introducing a dynamically rescaled time
variable, and its relation to the low-dimensional chaotic dynamics is
thus uncovered. The relation of this irregular motion with a strange
type of instability of heteroclinic cycles is also examined.

\end{abstract}

\pacs{keywords: nonlinear dynamics, chaos, heteroclinic cycle,
replicator, game dynamics, non-stationary motion, structural stability}

In this paper, we will report a new type of asymptotic behavior
observed in a game dynamics system. The behavior exhibits a combined
nature of the attraction to a heteroclinic cycle and a chaotic motion;
the trajectory visits some of the saddle points with an irregular
order and the lengths of the stay near the saddles are also irregular.
The irregularity here is related to an intrinsic chaotic dynamics and
not due to the randomness from noise. The behavior is robust against a
small variation of the parameters, and has a close relation with the
structurally stable heteroclinic cycles studied recently in various
contexts
{}\cite{Ashw-Swif}\cite{Knob-Silb}\cite{Hans-ETAL}\cite{Brannath}.

The game dynamics system we will study is a kind of population
dynamics, and is represented with a set of equations
\BEA
{d\over dt} x_i &=& \lambda_i x_i ,
\label{eqn:GAME1}\\
\lambda_i &=& s_i - \bar s ,
\label{eqn:GAME2}\\
s_i &=& \sum\limits_{j=1}^n g_{ij} x_j ,
\label{eqn:GAME3}\\
\bar s &=& \sum\limits_{i=1}^n s_i x_i ,
\label{eqn:GAME4}
\EEA
with a constraint,
\BE
\sum\limits_{i=1}^n x_i = 1,\qquad (0\le x_i\le 1).
\label{eqn:Constraint}
\EE
$x_i$ represents the renormalized population of species $i$, $s_i$ is
the score (or fitness) of species $i$, and $\bar s$ is the average
score over the population.  $\{g_{ij}\}$ defines the interaction
between species, each $g_{ij}$ representing the score of species $i$
in its battle with $j$.

Such a model was proposed by Taylor and Jonker in the context of the
evolution of strategy into ESS(evolutionary stable states)
{}\cite{Tayl-Jonk}; it is also regarded as a simplified model
equation for the molecular evolution \cite{Eigen}, and gives a minimal
model for the system of interacting self replicators \cite{Hofb-Sigm}.
It is also related to the Lotka-Volterra equation through a
transformation of variables \cite{Hofb-Sigm}.

The system exhibits various types of asymptotic behavior. The simplest
one is the relaxation to a resting state. The chaotic oscillation as
well as regular oscillation can be observed in some range of
parameters when 4 or more species are involved
{}\cite{Hofbauer}\cite{Gilpin}\cite{Arne-ETAL}. Attractive
heteroclinic cycles also are observed generically for systems with 3
or more species {}\cite{May-Leon}.

As is easily seen from the constraint (\ref{eqn:Constraint}), the phase
space of the game dynamics system with $n$ species is given by an
($n-1$)-dimensional simplex. The surface of the simplex consists of
$n$ hyper-planes $x_i=0$, each of which gives an invariant set.
Therefore, any intersection of these hyper-planes is also an
invariant set. The flow on such an invariant set is described by the
game dynamics equations of the original form but with a reduced number
of species, and gives a subsystem of the original one.
Note that there exist many ( at least $n$ ) fixed points on the
surface of the simplex, since each of these subsystems has at least
one fixed point.

A peculiar feature of this system is the existence of robust
heteroclinic orbits. This is basically due to the hierarchical
structure of the invariant set of the system described above. Suppose
that two saddles $A$,$B$ and a heteroclinic orbit from $A$ to $B$
exist on an invariant set which corresponds to a subsystem with $d$
species. The heteroclinic orbit will be robust against a small
variation of the parameters if
\BE
\dim A^u + \dim B^s > (d-1),
\EE
where $A^u$ denotes the unstable manifold of saddle $A$ in the
subsystem \cite{Guck-Holm}. Thus, if the saddle $B$ is an attractor in
the subsystem, i.e., $\dim B^s = (d-1)$, the heteroclinic orbit is
robust.  We will concentrate on such robust heteroclinic orbits,
therefore, every saddle which appears in this paper is an attractor in
the subsystem consisting of the species with non-zero $x_i$ at the
saddle point. Note that such saddles and robust heteroclinic orbits
are all on the border of the phase space, because it must belong to
a certain subsystem.

Some of the results presented in this paper are obtained from
numerical simulations. A small noise in the calculation may
drastically change the behavior of the trajectory especially in the
vicinity of the heteroclinic orbits. To suppress the numerical error,
the simulations were done with algorithm which keeps a constant
precision for them. It is realized by doing the numerical simulations
of the dynamics of $\log x_i$ which is, in principle, equivalent to
the original dynamics of $x_i$. The error for the population near the
zero value is drastically reduced by keeping tracks its logarithm. In
the contrast, the error for non-zero valued population is
automatically reduced by the dynamical contraction in the relevant
direction, as far as we work with robust heteroclinic orbits mentioned
above.

We will now look into the system behavior.
An example is presented in fig.1. It is obtained from the simulation
for a system with 5 species with interaction matrix
\begin{equation}
g_{ij}=\left (
\begin{array}{ccccc}
- -1.0& -20.0& -0.4& -1.0& 1.0\\
1.5& 0.0& -0.7& -7.3& 0.5\\
X& 1.0& 0.0& 0.0& -0.1\\
- -0.9& 0.8& 1.0& -1.0& -0.1\\
0.0& -8.0& 0.7& 1.3& 0.0
\end{array}
\right ),
\label{MATRIX}
\end{equation}
where $X=0.32$.  The population are set as $(1/5,1/5,1/5,1/5,1/5)$
initially.  The behavior looks like the attraction to an heteroclinic
cycle.  The trajectory visits several quasi-stable states, and the
period of the stay there shows a geometrical expansion. There is,
however, irregularity in the order of visits and lengths of the stays,
which is a major difference from typical cases of the attraction to an
heteroclinic cycle (fig.\ref{EXAMPLE2}).  The transition between these
two types of behavior, namely the irregular 'rambling' motion and the
regular attraction to heteroclinic cycle, will be studied afterwards
using the same matrix as the above but with different value of $X$.

We will analyze the asymptotic dynamics of this system. Here we assume
that the trajectory stays sufficiently long in the neighborhoods of
the saddles, while the transitions between them take place in a much
shorter time. The irregular rambling motion and also the attraction to
a heteroclinic cycle are analyzed as follows.  In the analysis, we
will work with the logarithm of the populations
\BE
y_i \equiv \log x_i,
\EE
and consider an approximation for the dynamics of $y_i$. Here
we write the approximated dynamics in the form
\BE
{d\over dt}Y_i =f_i(\vec Y),
\EE
where $Y_i$ is used instead of $y_i$ to avoid confusion and $\vec Y$
denotes $(Y_1,Y_2,\cdots,Y_n)$. It is assumed to be valid in the
vicinity of saddles and/or robust heteroclinic cycles.  All of those
relevant saddles and heteroclinic orbits are on the border of the
phase space by assumption. Therefore, some of the species have nearly
zero valued population in the considering region. We will call such
species as 'inactive' species and the others as 'active' ones.

The considering simplification consists of following two points,
namely,
\begin{itemize}
\item
ignoring the change of $Y_i$ made during the transition between saddles,
\item
linearizing the flow in the neighborhood of saddles.
\end{itemize}

The dominant contribution to the variation of $y_i$ comes from the
stays in the neighborhoods of saddles, if the transition between
saddles is sufficiently swift as assumed. Thus we will take an
approximation that the transition between saddles is an instantaneous
event, and neglect the change of $Y_i$ during the transition.  With
such a simplification, $Y_i$ contains non vanishing
error inevitably, and it is useless to consider a direct
correspondence between $Y_i$ and $x_i$. Thus we simply assume that
$Y_i$ is zero when species $i$ is active, and $Y_i<0$
corresponds to $x_i \approx 0$. In this case, the
constraint on $\vec Y$ corresponding to (\ref{eqn:Constraint}) is
represented as,
\BE
\max_i Y_i=0.
\EE

As is mentioned above, the transition between saddles is neglected in
the dynamics of $\vec Y$. Therefore, all of the change of $\vec Y$ is
made during the stay in the neighborhoods of certain saddles.
By linearizing the flow in the neighborhood of saddles, $\vec f$ can
be written formally as,
\BE
f_i(\vec Y)=\lambda_i^A,
\label{eqn:constraintY}
\EE
where $A$ denotes the neighboring saddle which should be determined
from $\vec Y$, and $\lambda_i^A$ is the growth rate of species $i$ at
saddle $A$.

Here we define $\vec f$ by simply assuming that $A$ is determined as
the attractor of the subsystem which consists of the species with
$Y_i=0$. Thus invasion of species $I$ corresponds to an instantaneous
event of $Y_I$ reaching 0, and immediately after that the system is
assumed to be in the neighborhood of a new saddle.  Note that this
definition of $\vec f$ may be invalid for some values of $\vec Y$,
where two or more attractor coexist in the associated subsystem.
This definition, however, is still valid for most value of $\vec Y$
and it is sufficient for the following argument.

Although $\lambda_i^A$ is zero if $x_i^A$ has nonzero value, $Y_i=0$
does not necessarily lead to ${\dot Y}_i=0$. This is because the
attractor can be on the border of the phase space of the associated
subsystem, that is, $x_i^A$ can be zero even if $Y_i$ is zero. In any
case, the constraint (\ref{eqn:constraintY}) is not violated because
$\lambda_i^A$ cannot be positive if $Y_i$ is zero, in order for the
fixed point $A$ to be an attractor in the associated subsystem.

Since the function $\vec f(\vec Y)$ depends only on whether each
component of $\vec Y$ is zero or not, it can be formally expressed as
a function of $\vec\eta \equiv \vec Y / |\vec Y|$.  Thus, by using the
notations
\BEA
\vec\eta &\equiv& {\vec Y\over |\vec Y|},\\
U &\equiv& |\vec Y|,\\
\vec g(\vec \eta) &\equiv&
\vec f(\vec\eta)-(\vec\eta\cdot\vec f(\vec\eta))\vec\eta,\\
h(\vec\eta) &\equiv& \vec\eta\cdot\vec f(\vec\eta),
\EEA
a set of equations,
\BEA
{d\over dt} \vec\eta &=& U^{-1}\vec g(\vec\eta),
\label{eqn:EtaT}\\
{d\over dt} U &=& h(\vec \eta),
\label{eqn:UT}
\EEA
is obtained. It should be noted that, corresponding to
(\ref{eqn:constraintY}), there is a constraint on $\vec\eta$, such
that,
\begin{equation}
\max\limits_i \eta_i = 0.
\label{eqn:constraintEta}
\end{equation}

We now introduce a dynamically rescaled time variable $\tau$ defined as
\begin{equation}
 {d\tau \over dt}= U^{-1}.
\label{eqn:TauT}
\end{equation}
Then the equations (\ref{eqn:EtaT}),(\ref{eqn:UT}) can be rewritten as
\begin{eqnarray}
{d\over d\tau} \vec \eta &=& \vec g(\vec \eta),
\label{eqn:Eta}\\
{d\over d\tau} (\log U) &=& h(\vec \eta).
\label{eqn:LogU}
\end{eqnarray}
Thus we obtain an autonomous equation for $\vec\eta$, and this determines
the sequence of the saddles visited in the rescaled time $\tau$. On the
other hand, the time scale of the motion which associated with the length
of the stay near saddles is determined by $U$. As can be seen from
(\ref{eqn:Eta}) and (\ref{eqn:LogU}), the motion has no
characteristic time scale.
Steady motion of $\vec\eta$ in $\tau$ results in a drift of $(\log U)$
with some average velocity measured with $\tau$. This results in a geometrical
change of the time scale of the motion in $t$ through (\ref{eqn:TauT}).
Thus a limit cycle solution of
(\ref{eqn:Eta}) corresponds to a heteroclinic cycle in the
original game dynamics system.

So far, it has not been questioned whether the trajectory of
$\vec\eta(\tau)$ is regular or chaotic. We now look into this point
more carefully with some case studies. For this purpose, it is
convenient to work with discrete maps which represent the change of
$\vec Y$ in a neighborhood of a saddle, rather than the continuous
flow of $\vec Y$ or $\vec \eta$.

The change of $\vec Y$ near a saddle is
represented by a piecewise-linear map. This is because when the
invader is specified as $I$-th species, the map is expressed in a
linear form,
\BE
Y_i^{(n)}\longmapsto Y_i^{(n+1)} =
Y_i^{(n)} - {\lambda_i^A\over \lambda_I^A}Y_I^{(n)},
\label{SADDLEMAP}
\EE
where $A$, $I$ denotes the saddle and invading species respectively,
and $\vec Y^{(n)}$ denotes the value of $\vec Y$ at the $n$-th
transition. The invader is determined in accordance with the value of
$\vec Y^{(n)}$, thus the map becomes 'piecewise'-linear as a whole.

The return map for $\vec Y$ at some Poincar\'e section can also be
represented by a piecewise-linear map. Actually, if the sequence of
saddles --- or equivalently the sequence of invaders ---
is specified, the return map for $\vec Y$ is obtained as a product of
the linear maps with the same form as (\ref{SADDLEMAP}). Thus
each segment of the return map is a linear map. The sequence of the
saddle is determined in accordance with the value of $\vec Y$ at the
cross section, and the return map also becomes piecewise-linear.

Let us consider the specific system (\ref{MATRIX}) and look into its
asymptotic behavior. A Poincar\'e section at the transition path from
saddle $\{3,4\}$ (where species 3 and 4 is steadily coexisting) to
saddle $\{5\}$ is considered.  The saddle $\{3,4\}$ is apart from the
hyperplanes $x_3=0$ and $x_4=0$, and after the invasion of species 5,
the orbit gets apart also from $x_5=0$.  Therefore, the value of $\vec
Y$ at the corresponding moment has only two non-zero components, $Y_1$
and $Y_2$. As is obvious from the above argument, the sequence of the
saddles visited remains unchanged if $\vec Y$ is multiplied by a
scalar factor. Thus the sequence of the saddles can be determined from
the ratio between these components, i.e.,
\BE
Z \equiv Y_1/Y_2.
\EE
In this way, we can analyze the saddle-sequence with the return map
for $Z$. If the chaotic trajectory is produced by this map, the
corresponding saddle-sequence will also be chaotic.

The return map for $\vec Y$ and also the return map for $Z$ can be
calculated analytically from the interaction matrix given in
(\ref{MATRIX}). The map for $Z$ is presented in fig.\ref{RETMAP1}. The
part of the map displayed consists of three elements, each of which
corresponds to a particular sequence of the saddles. The three types
of the sequence are presented in table \ref{TABLE}.  The map for $Z$
has one fixed point associated with heteroclinic cycle $C2$ which is
an unstable fixed point, and the map generates a chaotic sequence of
$C1$, $C2$ and $C3$ under general initial conditions.

The sequence of the saddles is thus expected to become chaotic in this
case, which is confirmed in the numerical simulation. A return map for
$z\equiv y_1/y_2$ is studied, and its Poincar\'e section is taken at
$x_5=\epsilon$, ${\dot x}_5>0$, $x_3>\epsilon$, $x_4>\epsilon$ with
$\epsilon=10^{-8}$. This hyper-plane corresponds to the instant of the
'invasion' of species 5 to the saddle $\{3,4\}$. The behavior of this
map is well reproduced by the return map for $Z$ which can be
calculated analytically, as is exhibited in fig.\ref{RETMAP1}. It is
thus clear that the irregularity of the sequence of the saddles
observed in figs.\ref{EXAMPLE1} and \ref{EXAMPLE2} is not due to noise
or numerical errors but generated by the intrinsic chaotic dynamics.

We have reduced the original 4 dimensional flow to a one dimensional
map. One of the variables is eliminated by taking a Poincar\'e section
of the flow. The reduction of the second variable is enabled by the
strong contraction of the flow, and we could separate the last one
because of the linear approximation. Note that it depends on the
particular system or particular cycle, that how many variables can be
reduced by the contraction of the flow. If all the relevant
heteroclinic orbits belong respectively to $1$ dimensional subsystems,
no variables can be reduced by the contraction.

Finally we will analyze the break down of a heteroclinic cycle
attractor, which leads to the emergence of the chaotic rambling
motion.

The return map for $Z$ (fig.\ref{RETMAP1}) has a fixed point
associated with heteroclinic cycle $C2$ as stated before. The
stability of this fixed point is related to the property of the matrix
$M^{(C2)}$ which represents the return map for $(Y_1,Y_2)$ associated
with the same cycle. The fixed point of the return map for $Z$
corresponds to an eigenvector of $M^{(C2)}$, and the condition for the
cycle $C2$ to be an attractor is expressed in terms of the eigenvalues
and eigenvectors of $M^{(C2)}$, that is,
\begin{itemize}
\item
$\Lambda_F > 1 $,
\item
$(\forall i\ne F),|\Lambda_F| > |\Lambda_i|$,
\end{itemize}
where $\Lambda_F$ is the eigenvalue associated with the fixed point of
the return map for $Z$.

Even if the first condition is not satisfied, the fixed point in the
return map remains stable. In this case, however, the iteration of
this cycle makes $|\vec Y|$ smaller and smaller, which means that the
trajectory diverges from the heteroclinic cycle gradually. On the
other hand, when the second condition is violated, the fixed point
becomes unstable. Since the map for $\vec Y$ is linear in the region
associated with $C2$, there is no nonlinear effect to suppress the
instability of the fixed point, so that the trajectory will inevitably
be thrown out of $C2$. In this way, the destabilization of the fixed
point in the return map leads directly to the loss of attractiveness
of heteroclinic cycle $C2$.

Matrix $M^{(C2)}$ can be calculated from the matrix in
(\ref{MATRIX}) as,
\BE
M^{(C2)}= \left (
\begin{array}{cc}
1.9389+6.418X & 1.2386-12.836X \\
2.841+8.42X   & 0.634-16.84X
\end{array}
\right ) .
\EE
The fixed point of the return map for $Z$ corresponds to the largest
eigenvalue of this matrix which is larger than 1 for the considering
range of $X$, so that cycle $C2$ is attractive if the fixed point of
the map is stable. The absolute value of the negative eigenvalue,
however, can exceed the largest eigenvalue, and destabilization of
the fixed point can be observed by changing parameter $X$. Its
critical value $X_c$ is obtained from the condition,
\BE
\hbox{Trace\ } M^{(C2)}=0,
\EE
and we obtain $X_c=0.24687\cdots$. If $X$ is larger than $X_c$, the
fixed point is unstable and the observed saddle-sequence becomes
chaotic. On the contrary, the attraction to the heteroclinic cycle
$C2$ is generic behavior when $X<X_c$. The second return map for $Z$
(fig.\ref{RETMAP2}) clearly shows that when the fixed point of the
return map becomes unstable, the cycle $C2$ ceases to be an attractor
and then a chaotic rambling motion appears.

The stability of the heteroclinic cycle has rather peculiar features,
as also noted by Melbourne \cite{Melbourne}.  Even in the case of
$X<X_c$, the heteroclinic cycle $C2$ is not stable in the usual sense,
because an arbitrarily small additive noise can kick the trajectory
out of $C2$ into a 'transverse' direction. It should also be noted
that even in the case of $X>X_c$, there exists a set of initial
conditions with zero measure corresponding to the unstable fixed point
in the map for $Z$, which is attracted to the cycle $C2$ (that is, the
'stable manifold' of $C2$). Another interesting phenomenon related to
stability of the structurally stable heteroclinic cycles is reported
recently \cite{Kirk-Silb}. That is the coexistence of two or more
heteroclinic cycle attractors sharing a common heteroclinic orbit. It
will correspond to the case when the return map for $\vec\eta$ has two
or more stable fixed points.

We have worked with the case of 5 species, because it is a minimal
system capable of displaying an irregular saddle-sequence as an
asymptotic behavior. Since $\vec\eta$ has two constraints,
$|\vec\eta|=1$ and $\max\limits_i \eta_i=0$, the dynamics of
$\vec\eta$ can be represented by a flow on a $n-2$ dimensional
spherical surface (with $n$ denoting the number of species).
Therefore the asymptotic behavior of $\vec\eta$ cannot be irregular if
$n$ is smaller than 5. It should be also noted that 5 species are
required also in Lotka-Volterra systems for the emergence of the
irregular rambling motions, if we assume that the total population
always has finite value.

We will mention the nature of the approximation used in this paper.
The trajectory of $\vec\eta$ gives a 'good' approximation for the
trajectory of $\vec y$, in the sense that for any positive integer $k$
and for generic value of the initial condition $\vec\eta_0$, there
exist a set of initial conditions of $\vec y$ which give the
saddle-sequences which have opening $k$ saddles in common with the one
associated with the trajectory of the $\vec\eta$.  It can be proven
with assuming the non-degeneracy of $\{g_{ij}\}$. A rough sketch of
the proof is given below. The complete proof will be given in another
paper \cite{Chawanya}.

We can confirm the existence of an upper limit for error associated
with the map (\ref{SADDLEMAP}) (where the error means the difference
between the change of $\vec Y$ made by this map and the change of
$\vec y$ made during the corresponding stay and transition) which does
not depend on the value of $\vec Y$. It is also confirmed that, as is
obvious from (\ref{SADDLEMAP}) and the assumption of non-degeneracy,
an upper limit exists for the magnification rate of the deviation of
two trajectories of $\vec Y$ in each step. Therefore, we can evaluate
the upper limit of the difference between the value of $\vec y$ and
its approximation $\vec Y$ after $k$ transitions.  With considering
the fact that the map (\ref{SADDLEMAP}) is linear, we can choose the
initial condition $\vec Y_0$, s.t. i) the corresponding
saddle-sequence is the same as the one given by the trajectory started
from $\vec\eta_0$, and ii) the opening $k$ saddles of the associated
saddle-sequence is not changed even if the error is taken into
account. $\vec Y_0=U_0\vec\eta$ with a sufficiently large $U_0$ gives
an example of the 'good' initial condition for $\vec Y$, and the
trajectory with the initial condition for $\vec y$ corresponding to
the $\vec Y_0$ gives the saddle-sequence with the desired property.

As is mentioned above, we have analyzed the global behavior of the
trajectory basically in the logarithmically transformed coordinate.
Such a treatment is enabled by the robust existence of the
hierarchically structured invariant sets ($\prod x_i=0$, which form
the border of the phase space), and does not rely on a specific form
of interaction.  Thus the same method can be applied to the analysis
of the system with nonlinear interaction as well. As is reported by
several authors, structurally stable heteroclinic cycles can exist in
systems with a certain kind of symmetry. Such systems have invariant
sets with similar structure to that in the game dynamics system. Thus
the chaotic rambling over quasi stable states could also be observed
in such systems.

The relation between $\vec Y$ and $\vec f$ is not simple in the
present system, and it is difficult to write down explicitly the
rescaled dynamics (\ref{eqn:Eta}). The dynamics of the transitions
among saddles can be treated more elegantly in the case of a coupled
logistic equations, where the dynamics with a rescaled time is written
explicitly, and the emergence of a chaotic saddle-sequence is
demonstrated \cite{Sasa-Chaw}.

The author is grateful to S. Sasa for extensive discussion and
collaboration in related works. He also thanks Y.Kuramoto for
enlightening advice, and T.Ikegami, H.Tanaka, T.Yanagida, K.Kaneko,
and K.Okuda for valuable suggestions. T.Shichijo and K.Watanabe are
acknowledged for collaboration in an earlier stage of this study, and
G.C. Paquette for the careful reading of the earlier version of this
manuscript. Finally referee's thoughtful comments are appreciated.
This work was supported by the Yukawa foundation.

\vskip1cm

\begin{table}
\begin{tabular}{|c|lc|} \hline
 $Z$&{sequence of the saddles \hfill}& \\ \hline
 (0.791,0.902)& $\{3, 4\}\rightarrow \{5\}\rightarrow\{1,
5\}\rightarrow\{2\}\rightarrow\{3\}\rightarrow\{5\}\rightarrow\{2\}\rightarrow\{3\}\rightarrow\{3, 4\}$&:C1\\
 (0.902,2.0)& $\{3, 4\}\rightarrow \{5\}\rightarrow\{1,
5\}\rightarrow\{2\}\rightarrow\{3\}\rightarrow\{3, 4\}$&:C2\\
 (2.0,$\infty$)& $\{3, 4\}\rightarrow
\{5\}\rightarrow\{2\}\rightarrow\{3\}\rightarrow\{3, 4\}$&:C3\\
\hline
\end{tabular}
\caption{Relation between the value of $Z$ and the sequence of saddles.}
\label{TABLE}
\end{table}


\begin{figure}
\caption{An example of the temporal evolution of the population of
species, in the case of attractive heteroclinic network.
The population of all 5 species are plotted.  $\{g_{ij}\}$ is
chosen as in (\protect\ref{MATRIX}) with $X=0.32$, and the initial condition
is set as $x_1=x_2=\cdots=0.2$. Numerical simulation is done with the
Runge-Kutta method with variable step size, taking $\log{}x_i$ as
variables. }
\label{EXAMPLE1}
\end{figure}


\begin{figure}
\caption{A temporal sequence of the saddles for a regular
({\bf a:}$X=0.24$), and an irregular motion ({\bf b:}$X=0.32$).  The
abscissa represents time in the log scale, and the ordinate
displays the nearest saddle.  The transition between saddles is
omitted in this figure, however, the process is so relatively fast
that it constitutes practically zero width with this scale.  The
initial condition is the same as in fig.1.  }
\label{EXAMPLE2}
\end{figure}


\begin{figure}
\caption{First return map for $Z$ (solid line; analytically
calculated), and $z\equiv y_2/y_1$ (plotted with circles; obtained
from a numerical simulation). The parameter $X$ is set as $X=0.32$,
and the considered Poincar\'e section corresponds to the exit from
saddle \{3,4\} towards saddle \{5\}.  The section is taken as
$x_5=\epsilon$, ${\dot x}_5>0$, ($x_3 > \epsilon$, $x_4 > \epsilon$)
with $\epsilon=10^{-8}$ in the numerical simulation.  Plotted data
corresponds to successive 200 turns (corresponding range of $\log t$
is $[27.5,105.25]$).  }
\label{RETMAP1}
\end{figure}


\begin{figure}
\caption{The second return map for $Z$, obtained for $X=0.24$ (dashed line)
and 0.25 (solid line). The considered section is the same as in fig.3.
}
\label{RETMAP2}
\end{figure}

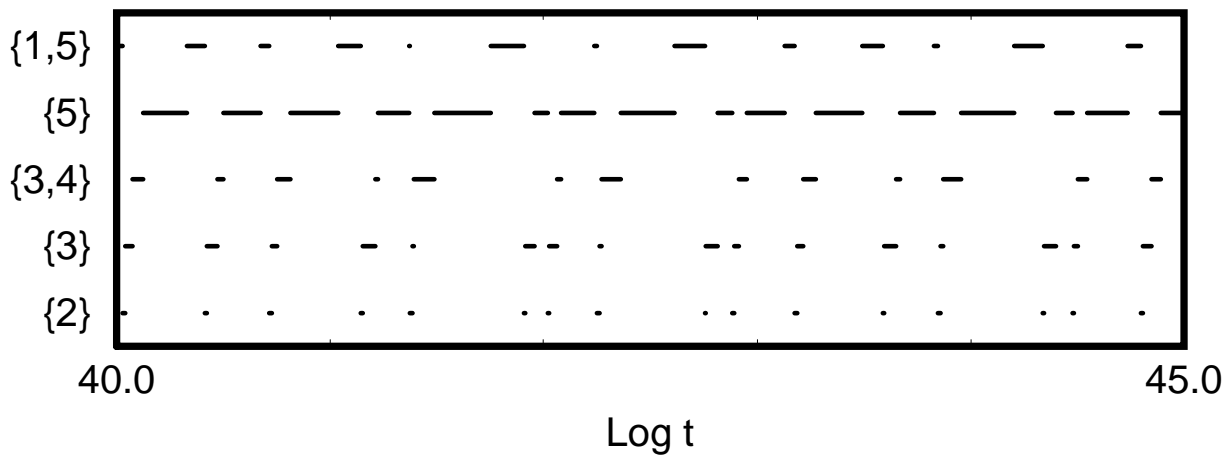
\end{document}